\documentclass[fleqn,usenatbib]{mnras}

\usepackage{newtxtext,newtxmath}

\usepackage[T1]{fontenc}

\DeclareRobustCommand{\VAN}[3]{#2}
\let\VANthebibliography\thebibliography
\def\thebibliography{\DeclareRobustCommand{\VAN}[3]{##3}\VANthebibliography}

\usepackage{graphicx}	
\usepackage{amsmath}	
\usepackage{comment}
\usepackage{physics}
\usepackage{bm}

\usepackage{url}

\usepackage{booktabs}
\usepackage{multirow}

\usepackage[hang,small,bf]{caption}
\usepackage[subrefformat=parens]{subcaption}
\usepackage{sepfootnotes}

\title[3D-Forecasting SN explosions using deep learning]{3D-Spatiotemporal Forecasting the Expansion of Supernova Shells Using Deep Learning toward High-Resolution Galaxy Simulations}

\author[K. Hirashima et al.]{Keiya Hirashima$^{1}$\thanks{JSPS Research Fellow }\thanks{E-mail: hirashima.keiya@astron.s.u-tokyo.ac.jp},
Kana Moriwaki$^{2,3}$,
Michiko S. Fujii$^{1}$,
Yutaka Hirai$^{4,5}$\footnotemark[1],
Takayuki R. Saitoh$^{6}$, and
\newauthor Junichiro Makino$^{6}$
\\
$^1$Department of Astronomy, Graduate School of Science, The University of Tokyo, 7-3-1 Hongo, Bunkyo-ku, Tokyo 113-0033, Japan \\
$^2$Department of Physics, Graduate School of Science, The University of Tokyo, 7-3-1 Hongo, Bunkyo-ku, Tokyo 113-0033, Japan\\
$^3$Research Center for the Early Universe, The University of Tokyo, 7-3-1 Hongo, Bunkyo, Tokyo 113-0033, Japan\\
$^{4}$Department of Physics and Astronomy, University of Notre Dame, 225 Nieuwland Science Hall, Notre Dame, IN 46556, USA\\
$^{5}$Astronomical Institute, Tohoku University, 6-3 Aoba, Aramaki, Aoba-ku, Sendai, Miyagi 980-8578, Japan \\
$^{6}$Department of Planetology, Graduate School of Science, Kobe University, 1-1 Rokkodai-cho, Nada-ku, Kobe, Hyogo 657-8501, Japan
}

\date{Accepted XXX. Received YYY; in original form ZZZ}

\pubyear{2023}

\graphicspath{{./Figures/}}
\begin{document}
\label{firstpage}
\pagerange{\pageref{firstpage}--\pageref{lastpage}}
\maketitle

\begin{abstract}
Supernova (SN) plays an important role in galaxy formation and evolution. In high-resolution galaxy simulations using massively parallel computing, short integration timesteps for SNe are serious bottlenecks. This is an urgent issue that needs to be resolved for future higher-resolution galaxy simulations. One possible solution would be to use the Hamiltonian splitting method, in which regions requiring short timesteps are integrated separately from the entire system. To apply this method to the particles affected by SNe in a smoothed-particle hydrodynamics simulation, we need to detect the shape of the shell on and within which such SN-affected particles 
reside during the subsequent global step in advance. In this paper, we develop a deep learning model, 3D-MIM, to predict a shell expansion after a SN explosion. Trained on turbulent cloud simulations with particle mass $m_{\rm gas}=1$ M$_\odot$, the model accurately reproduces the anisotropic shell shape, where densities decrease by over 10 per cent by the explosion. We also demonstrate that the model properly predicts the shell radius in the uniform medium beyond the training dataset of inhomogeneous turbulent clouds. We conclude that our model enables the forecast of the shell and its interior where SN-affected particles will be present.
\end{abstract}




\section{Introduction}
\label{sec:intro}

Supernova (SN) explosions are energetic events at the end of stars that cause a massive release of energy. The ambient interstellar medium (ISM) is heated up and swept away by the explosion. The resulting galactic outflow and turbulence are known to affect the star formation rates and the scale heights of galaxies \citep[][for review]{Naab+17}.
SNe can significantly impact galaxy formation and evolution along with other processes, including gravitational and hydrodynamic forces, radiative cooling and heating, star formation, and chemical evolution. 
As these processes complicatedly interact with each other, galaxy formation is commonly studied using numerical methods.
To accurately model galaxy formation, it is essential to properly incorporate the effects of the SN explosions into simulations.

Thanks to the development of supercomputers, recent Milky Way (MW) sized galaxy formation simulations have improved spatial and mass resolutions. Even in such simulations, the SN effects are dealt with in sub-grid models.
For instance, the `Latte' site of FIRE-2 simulation \citep{Hopkins+2018} for a MW-sized halo adopts $~ 7.0 \times 10^3~\rm M_{\odot}$ for an initial gas-particle mass.
The LATTE simulations use a mixed feedback injection scheme,
where they distribute thermal energy when the SN-explosion is resolved and use momentum feedback otherwise.
The `Mint' resolution of DC Justice League simulation \citep{Applebaum+2021}
achieves a mass resolution of $\sim$$10^3~\rm M_\odot$ for gas and star in MW-like zoom-in cosmological simulations
using an analytic model \citep{Stinson+2006} that distributes the mass, metals, and energy from SNe into surrounding gas particles.
Auriga simulation \citep{Grand+2021} achieves a mass resolution of $800~\rm M_\odot$ for gas in forming the satellite galaxies in a MW-mass halo in a cosmological zoom-in simulation.
They use mechanical feedback \citep{Vogelsberger+2013} to model the galactic wind and outflows driven by SNe.

If we go higher resolution, SNe should be computed without sub-grid models.
Previous studies on detailed blastwave simulations have shown that a finer mass resolution than $\sim$~1 M$_\odot$ is required to resolve the hot-phase build-up in a Sedov-Taylar phase \citep{Kim+2015, Hu+2016, Steinwandel+2020}.
While even state-of-the-art MW-sized simulations have not reached this resolution yet, some successful examples exist in those of relatively small galaxies.
The LYRA simulation \citep[][]{Gutcke+2022}
achieves $4~{\rm M_\odot}$ gas-mass resolution for a dwarf galaxy with a halo mass of $~ 2 \times 10^9~{\rm M_\odot}$ in a cosmological simulation to $z=0$ and is able to capture the propagation of SN blast waves.
For an isolated dwarf galaxy simulation, even higher resolutions are achieved: 
$4~{\rm M_\odot}$ \citep{Hislop+2022}
and $\sim$$1~{\rm M_\odot}$ \citep{Hu+2019} for the gas mass.
The SIRIUS simulation \citep{Hirai+2021}
achieves a gas-mass resolution of
$18.5 ~\rm M_\odot$ in cosmological zoom-in simulations of ultrafaint dwarf galaxies.

Computational issues remain to be addressed to achieve MW-sized simulations with $\sim 1~\rm M_\odot$ resolution.
To conduct such simulations, we need to use more than $10^{10}$ gas and star particles, at least two orders of magnitude larger than those for the current similar simulations.
One of the most serious bottlenecks is the short timesteps required for resolving blast waves in dense regions.
In recent high-resolution galaxy simulations, such as the dwarf simulations described above, individual (hierarchical) timesteps are employed to decrease the number of redundant calculations. 
The SN-affected particles, which comprise a tiny fraction of the entire galaxy, are integrated using much shorter timesteps than the rest of the particles, for which a large and shared timestep (global timestep) can be employed.
However, even with this method, in MW star-by-star simulations, the problem of communication overhead would still remain; the data transfer among nodes is required at every smallest timestep, and a large amount of CPU cores are forced to idle during such communications.\footnote{The overhead of communication becomes more dominant and worsens the scalability as the number of processors and/or steps increases. In recent particle-based galaxy-formation simulations, the scaling saturates on $\sim 10^3-10^4$ CPU cores \citep{Hopkins+2018,Springel+2021}.
}
One solution to this is to send SN-affected particles to a separate set of CPUs and integrate small-scale phenomena separately from the entire system during a global timestep with techniques such as the Hamiltonian splitting method \citep{Wisdom+1991, Saha+1994, Xu1995, Fujii+2007, Jaenes+2014, Springel2005,Springel+2021,Ishiyama+2009, SaitohMakino2010, Pelupessy+2012, Rantala+2021, Rantala+2022}.

Currently, the application of such a splitting method for SN feedback is not developed because it is difficult to predict the expansion of the SN shells and to pre-identify the particles that should be separately computed.
The simplest approach is using an analytic solution for the time evolution of the SN shell in an isotropic and uniform interstellar medium (ISM) \citep{Sedov1959}. However, since the ISM of the actual galactic environment is neither isotropic nor uniform, the SN shell would expand anisotropically, and this method fails to accurately predict the expansion regions of the SN shell in galaxy simulations. 
We need a more sophisticated method that can capture the complex expansion.

In this study, we propose to use deep learning to forecast the expansion of the SN shell.
In astronomy, deep learning has been used to solve many problems.
In particular, convolutional neural networks (CNNs) are commonly used for analysing spatial features of 2D or 3D data obtained in simulations;
2D CNNs have been used to emulate the dynamics of baryonic simulations \citep[e.g. ][]{Chardin+2019, Duarte+2022},
and 3D CNNs have been used
to increase the resolution of cosmological simulations \citep[e.g.][]{Li+2021} and emulate turbulent hydrodynamical simulations \citep[][]{Chan+2022}.
Recurrent models extract temporal trends in sequential data and are also used in astronomy.
For instance, \citet[][]{Lin+2021} used a recurrent neural network (RNN) to analyse gravitational wave signals.
Combining a CNN and recurrent model will allow us to analyse the time series of spatial distribution data obtained in a simulation.
In this study, we develop a deep learning model composed of 3D CNNs and recurrent modules to predict the 3D expansion of SN shells in inhomogeneous ISM.

We study deep learning forecasts of SN shells in dense regions 0.1 Myr after the explosion.
In this paper, we will show that our model can predict the evolution of SN shells, which contain SN particles inside, in the turbulent ISM with gravity and radiative cooling and heading.
The paper is organized as follows.
In Section \ref{sec:DataPre}, we explain the setup of our SN simulations and the data preparation methodology. 
We describe the models to the SN shells in Section \ref{sec:Methods}, where Section \ref{sec:method_analytic} lays down the analytic solution and Section \ref{sec:method_CV} provides our new deep learning model.
In Section \ref{sec:result}, we present the results of forecasting the expansion of SN shells using our deep learning model and evaluate the generalization performance.
In Section \ref{sec:discussion}, we discuss practical implementations of our new model in future galaxy simulations. Section \ref{sec:particle-selection} demonstrates how our approach can be utilized to detect SN-affected particles. Section \ref{sec:OtherPhysics} also explores potential ways to integrate our model into upcoming simulations of galaxies.
Finally, Section \ref{sec:conclusion} summarizes the paper and discusses the potential extension of our model and implementation of our future high-resolution galaxy simulations.

\section{Data Preparation}
\label{sec:DataPre}
We prepare training data from SPH simulations of a SN explosion inside the inhomogeneous density distribution of molecular cloud for deep learning.
The resolution of these simulations is adjusted to our upcoming star-by-star galaxy simulations. 

\subsection{Code}
\label{sec:code}

We simulate SN explosions using our $N$-body/SPH codes, \textsc{ASURA-FDPS} \citep{Saitoh+2008,Iwasawa+2016} and \textsc{ASURA+BRIDGE} \citep[][]{Fujii+2021}.
In these codes, hydrodynamics are calculated using the density-independent SPH (\textsc{DISPH}) method \citep[][]{Saitoh+2013, Saitoh+2016}, which can properly estimate the density and pressure at the contact discontinuity and reproduce fluid instabilities.
The gas properties are smoothed using the Wendland $C^4$ kernel \citep[][]{2012+Dehnen}.
The kernel size is set to keep the number of neighbour particles at $125\pm1$.
We adopt the shared timesteps with the Courant-like hydrodynamical timestep based on the signal velocity \citep[][]{Monaghan1997, Springel2005}.
The leap-frog method is used for time integration.
The timestep of a particle $i$ is written as
\begin{equation}
    \Delta t_i = C_{\rm CFL} \frac{2 h_i}{\max_{j} [c_i + c_j - 3w_{ij}]}
    \label{eq:CFL}
\end{equation}
where $C_{\rm CFL}=0.3$, $h_i$ and $c_i$ are the SPH kernel length and sound speed of a particle $i$, and 
\begin{align}
    w_{ij} & = {\bm v}_{ij} \cdot {\bm r}_{ij}/r_{ij},
\end{align}
where ${\bm r}_{ij}$ and ${\bm v}_{ij}$ are the relative position and velocity between particles $i$ and $j$, respectively. 
The index $j$ runs over all the particles within the kernel size from the particle $i$.
The timestep for integration $\Delta t$ is then determined as
\begin{equation}
   \Delta t = \min_{i} \Delta t_i.
\end{equation}
We adopt the artificial viscosity term originally introduced by \citet[][]{Monaghan1997}.
The time-dependent variable viscosity parameter form proposed by \cite{Rosswog2009} is also used.
The range of the viscosity parameter is 0.1 -- 3.0 in our simulations.
We use the metallicity-dependent cooling and heating functions from
$10$ to $10^9$ K generated by \textsc{Cloudy} version 13.5 \citep[][]{Ferland+1998, Ferland+2013,Ferland+2017}.
Assuming the environment of the Milky Way Galaxy, metallicity of $Z=1Z_\odot$ is adopted in our runs.

\subsection{Supernova Simulation Setup}
\label{sec:SPHSim}
We design the simulations as a SN explosion in a high-density star-forming molecular cloud with a large density contrast.
We assume an adiabatic compression of a monatomic ideal gas, which follows the equation of state with the specific heat ratio $\gamma=5/3$:
\begin{equation}
    P=(\gamma-1) \rho u,
\end{equation}
where $P$, $\rho$, and $u$ are the pressure, smoothed density, and specific internal energy, respectively.
The adiabatic compressible gas clouds follow the following equations:
\begin{align}
    \frac{d \rho}{dt} & = -\rho \nabla \cdot \bm{v}, \\
    \frac{d^2 \bm{r}}{dt^2} & = -\frac{\nabla P}{\rho} + \bm{a}_{\rm visc}-\nabla \Phi, \\
    \frac{d u}{dt} & = -\frac{P}{\rho} \nabla \cdot \bm{v} + \frac{\Gamma-\Lambda}{\rho},
\end{align}
where $r$ is the position, $a_{\rm visc}$ is the acceleration generated by the viscosity, $\Phi$ is the gravitational potential, $\Gamma$ is the radiative heat influx per unit volume, and $\Lambda$ is the radiative heat outflux per unit volume.

The process of the SN simulations is as follows.
First, to make gas clouds similar to star-forming regions, we evolve initially uniform-density gas spheres with a turbulent velocity field that follows $\propto v^{-4}$ for one initial free-fall time.
As for the initial gas sphere of the fiducial dataset, a total mass of $10^6~ \rm M_\odot$, a radius of 60 pc, and a uniform density of 190~cm$^{-3}$ are adopted.
The initial conditions are constructed using the Astrophysical Multi-purpose Software Environment \citep[][]{Portegies+2013,Pelupessy+2013,Portegies+2018}.
We use SPH particles with a mass of 1 $\rm M_\odot$ and an initial temperature of 100 K. 
These parameters are chosen to be comparable to the galaxy simulations we aim to perform.
We confirm that the mass resolution of our SN simulations is sufficient to reproduce the blast wave properties of SNe 
by performing the Sedov blast wave test \citep[][]{Sedov1959,Saitoh+2013} (see Appendix \ref{sec:blastwave}).
The softening length for gravitational interactions is set to 0.9 pc.

\begin{figure*}
    \includegraphics[width=2\columnwidth]{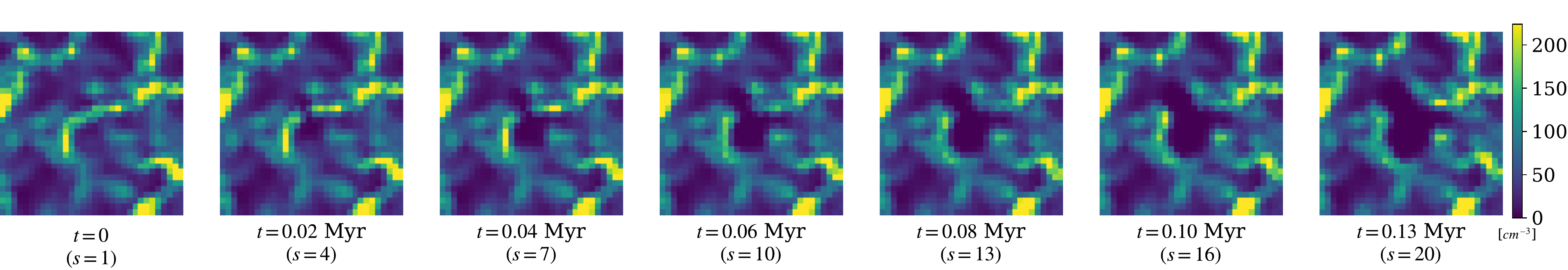}
    \caption{
    Snapshots of our SN simulation of density distribution for the Fiducial data. Figures show the time variation of density after a SN explosion until 0.133 Myr. 
    }
    \label{fig:ValTrend}
\end{figure*}

Then, the thermal energy of $10^{51}$ erg is injected into 100 SPH particles in the centre of the turbulent gas clouds as the explosion energy. The thermal energy is distributed to the particles by smoothing with a SPH kernel. SNe happen at the centre of the turbulent gas clouds in the simulations. The simulations of a SN explosion are performed for up to 0.2 Myr from the beginning of the energy injection. 
The upper panels of Fig. \ref{fig:ValTrend} show an example of the evolution of the gas density distribution after a SN explosion.

We also conduct simulations with a shorter duration and denser initial conditions. 
In addition, we consider two extra cases with different mean densities where there is no density fluctuation. 
All these simulations were used to generate training and test datasets.
The dataset names and parameters are summarized in Table~\ref{tab:datasets_table}.
Our simulations were performed using the supercomputer Cray XC50 at the Center for Computational Astrophysics (CfCA), National Astronomical Observatory of Japan.

\begin{table*}
	\centering
	\caption{The simulation properties (i.e. the average density $\bar{\rho}$, spatial symmetries of gas clouds, and the forecast horizon $t$) of the datasets used for training and testing our deep learning models.
     We have four datasets: Fiducial, Short-Term, Uniform, and Dense-Uniform.
     We use the Fiducial and Short-Term datasets to train deep learning models.
	The combinations of datasets used for training and testing are shown in Table \ref{tab:experiments_table}.
	}
	\label{tab:datasets_table}
	\begin{tabular}{lrcr} 
		\hline
		  Datasets   & $\overline{\rho}$ [cm$^{-3}$]   & Spatial Symmetries &  $t$ [Myr]\\
		\hline
		Fiducial & $4.1 \times 10  - 6.5 \times 10 $   & non-uniform \& anisotropic &  0.133\\
		Short-Term &  $4.1 \times 10  - 6.5 \times 10 $    & non-uniform \& anisotropic &  0.066\\
		\hline
		\hline
		Uniform & $5.6 \times 10$   & uniform \& isotropic &  0.133\\
        Dense-Uniform & $1.9 \times 10^2$   & uniform \& isotropic &  0.133\\
		\hline
	\end{tabular}
\end{table*}

\subsection{Training Data for Deep Learning}

For deep learning, we extract the central cubic regions with a side length of $60~\rm pc$
and obtained 3D volume images composed of $32^3$ voxels 
with a spatial resolution of 1.9 pc by smoothing gas particles with SPH kernels of size depending on the local densities.
The SNe are placed at the centre of the 3D volume images.
The extracted regions are small enough compared to the turbulent clouds, and the boundary effects can be ignored.
We performed 300 independent simulations of supernova explosions 
with different seeds for the turbulence velocity fields and got 20 snapshots with a time step $\Delta t$.
The initial turbulent gas clouds for the fiducial training datasets have mean densities ranging from $4.1 \times 10 ~ {\rm cm^{-3}}$ to $~ 6.5 \times 10 ~ {\rm cm^{-3}}$. The average and standard deviation of the mean densities are $5.4\times 10 ~ {\rm cm^{-3}}$ and 4.0 ${\rm cm^{-3}}$, respectively. In training, the densities are normalized by $1.9\times10^2$ ${\rm cm^{-3}}$ so that the majority of the data values fall within the range of 0 to 1.

We prepare two training datasets with different time steps as listed in Table \ref{tab:datasets_table}:
$\Delta t = 7.0 \times 10^{-3}$ Myr (Fiducial)
and $\Delta t = 3.5 \times 10^{-3}$ Myr (Short-Term dataset).
A sequence of data converted from each simulation contains 20 volume images.
We call a volume image in a sequence a frame.

To augment training data, flipped volume images and images with swapped axes (transposition) are added to the training data.
There are $2^3=8$ possible combinations of axes to flip and $3!=6$ combinations of axes to swap.
We thus generate $48 (=8\times 6)$ data sequences 
for each of the 300 simulations.

\section{Methods to Predict SN Shell Explosion}
\label{sec:Methods}
\subsection{Analytic Solution of the Expansion of Supernova Shell}
\label{sec:method_analytic}

The expansion of the supernova shell is described as a self-similar solution known as the Sedov-Taylor solution \citep{Sedov1959}.
This solution approximates the SN as a spherical point explosion in a uniform medium.
Introducing the dimensionless similarity variable $\xi$, the time evolution of the radius of the SN shell is written as 
\begin{equation}
    R(t) = \xi \left(\frac{E}{\rho}\right)^{1/5}t^{2/5},
    \label{eq:Sedov}
\end{equation}
where $E$ and $\rho$ are the energy injected by SN and the density of the surrounding ISM, respectively.

Equation (\ref{eq:Sedov}) is the analytic solution that can be used under the ideal environment of uniform density. However, the interstellar gas is not uniform, and therefore this equation is not always applicable for predicting SN shell evolution. We thus need to develop a model for non-uniform environments.

\subsection{Prediction for the SN Explosion}
\label{sec:method_CV}


\begin{figure*}
	\includegraphics[width=1.8\columnwidth]{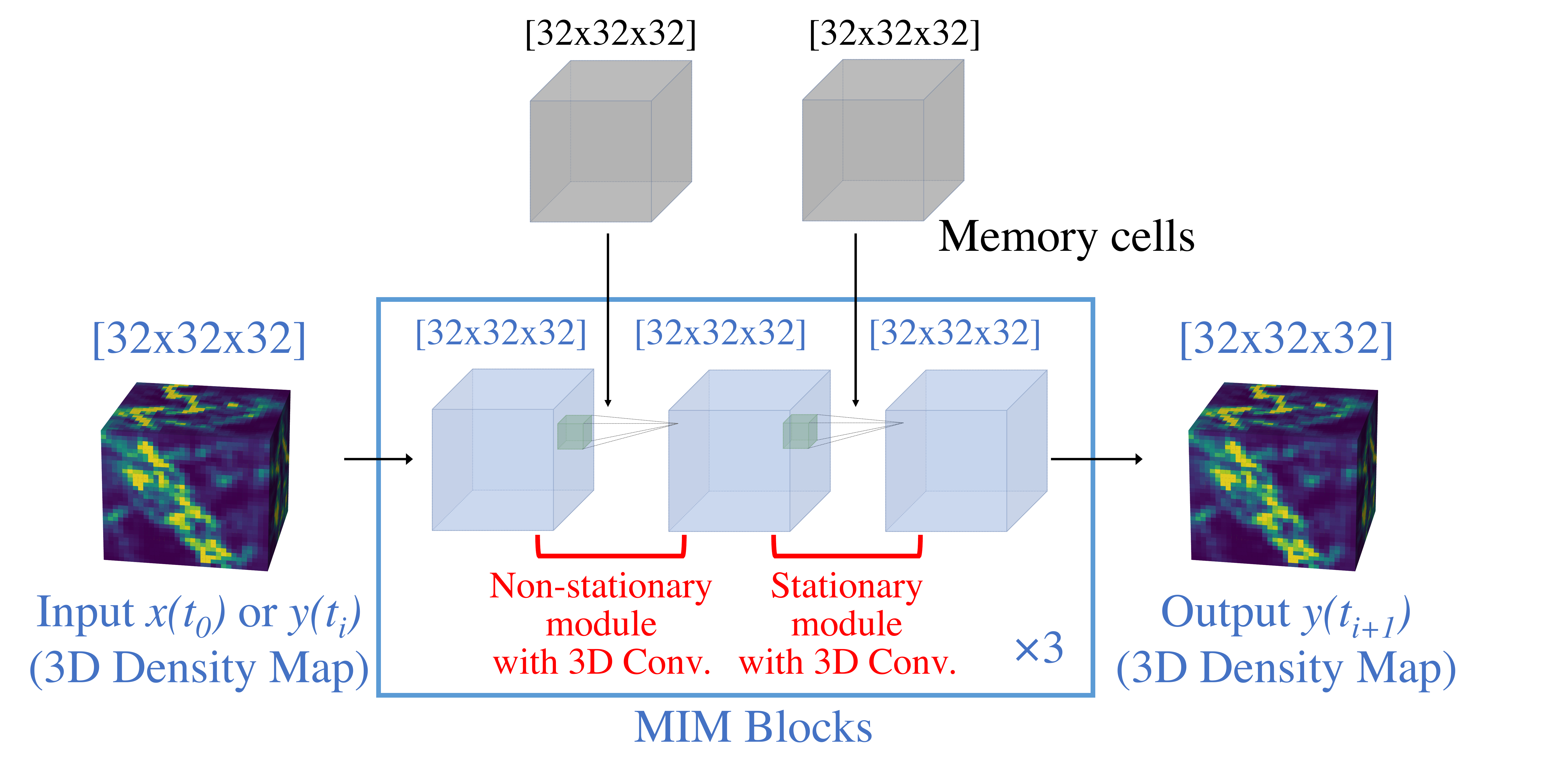}
    \caption{
    The architecture of our deep learning framework, 3D-MIM, which consists of three MIM blocks.
    The model predicts 3D gas density distributions at a timestep $t_{i+1}$ from those at the previous timestep $t_i$.
    A MIM block includes two modules to capture non-stationary and stationary variations,
    where the input data are convolved with memory cells.
    We omit the detailed internal structures of the modules, 
    see \citet{Wang+2018} for more details.
    }
    \label{fig:Model}
\end{figure*}

Several deep learning techniques for video prediction have been developed in the past several years. 
In particular, deep learning models combining CNNs and RNNs,
which learn spatial correlations in images and 
temporal correlations in time series, 
are often used \citep[for review, ][]{Oprea+2020}.
To forecast the time series data of the SN explosion,
we develop such a combined model based on \textsc{Memory In Memory (MIM) network}\footnote{\url{https://github.com/Yunbo426/MIM}} developed by \cite{Wang+2018}.

The schematic diagram of our model is shown in Fig. \ref{fig:Model}.
The model predicts 3D gas density distributions at a timestep $t_{i}$ $(i = 1,2,...,20)$ from those at the previous timestep $t_{i-1}$, where $t_{i} - t_{i-1} = \Delta t$.
We denote the simulated and predicted density distributions at time $t$ as $\bm{x}(t)$ and $\bm{y}(t)$, respectively.
The input data are either $\bm{x}(t_0)$ or $\bm{y}(t_i)$ $(i=1,2,...19)$.
By repeating the next-timestep predictions 20 times from $t_0 = 0$, 
we forecast the gas density distribution at $t_{20} = 20\Delta t$.

Our model has three MIM blocks with two recurrent modules with convolutional operations.
In each module in the MIM block, the input data are convolved with a set of 3D data cubes called memory cells.
The memory cells are initialized at the first prediction and are updated at each prediction, storing the information on the time variation of the gas distribution over all previous timesteps. 
We initialize the memory cells with a uniform distribution. 
The two modules are designed differently to capture stationary and non-stationary variations in data, respectively,
allowing the model to learn complex, non-stationary evolution of SN explosions more efficiently 
than conventional methods.\footnote{
Traditional time-series forecasting models 
have been developed for stationary time series, which have a constant mean and autocovariance over time \citep[for review, ][]{Wilson+2016}.
In such models, non-stationary time series is transformed into stationary data in some ways (e.g. linear and logarithmic transformations) and then predicted. However, high-order non-stationarity cannot be captured with such an implementation \citep[see discussion in][]{Wang+2018}.
}
See \citet{Wang+2018} for more details on the inner architecture of the modules.

Our new implementation for the original MIM network is the following two points.
First, we increase the internal dimension of the model by one, from 2D to 3D,  
to deal with the 3D density distribution.
Second, we change the number of input and output data cubes. 
The original MIM network is a many-to-many forecast model  
predicting the last ten frames from the initial ten.
We design our model as a one-to-many forecast model that takes an image just before the SN explosion as an input and returns 
all the gas distributions of subsequent timesteps.
We call our new forecast model \textsc{3D-MIM}\footnote{\url{https://github.com/kyaFUK/3D-MIM}}.

We implement our model using TensorFlow\footnote{\url{https://github.com/tensorflow/tensorflow}}
and train it using the NVIDIA GeForce RTX 3090 GPU installed on the desktop computer of the corresponding author and the NVIDIA A100.
The model is trained with L2 loss function between the true (simulated) and predicted gas density distributions for 64 epochs with batch size 4.
We adopt ADAM optimizer \citep{Kingma+2014} with a learning rate of 0.001.
For inference, computationally expensive architectures in our deep learning models (e.g. 3D-CNNs) are optimized to the supercomputer Fugaku by \textsc{SoftNeuro}\textregistered~\citep[][]{Hilaga+2021} developed by Morpho, Inc.

\section{Result}
\label{sec:result}

\begin{figure*}
 \includegraphics[width=2\columnwidth]{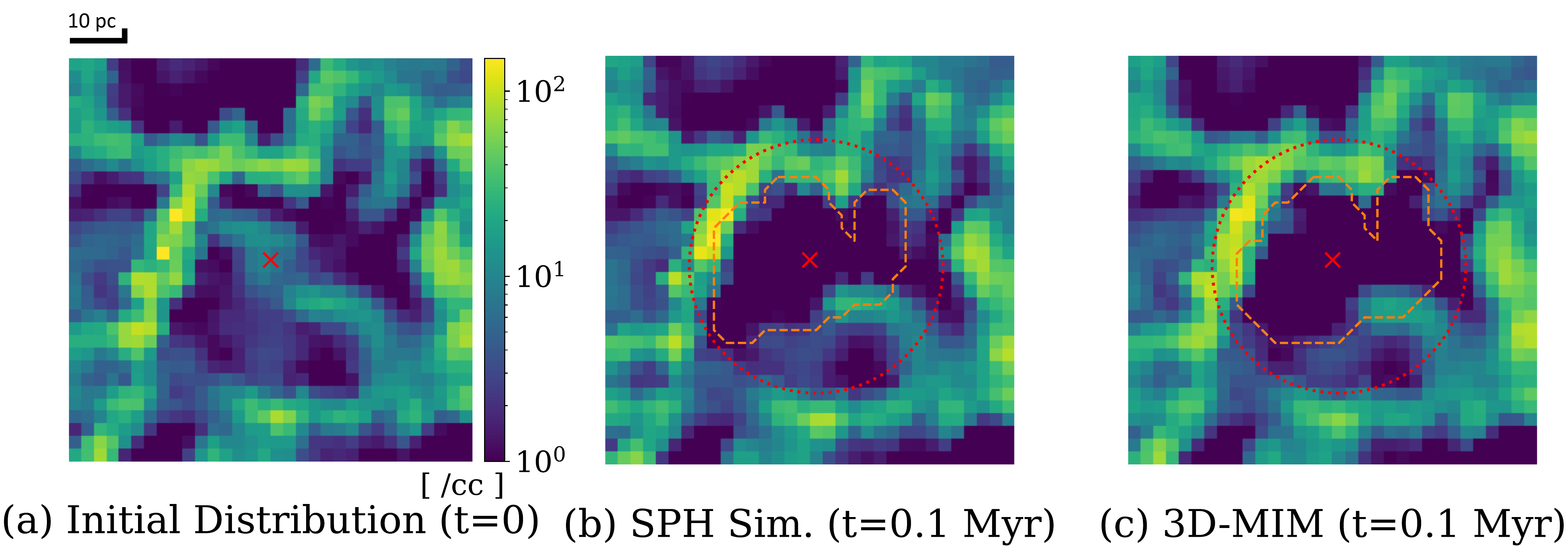}
    \caption{A forecast result by the 3D-MIM for Experiment 1 in Table \ref{tab:experiments_table}.
    Panel (a): the initial distribution just before the supernova explosion at the centre. Panel (b): the simulation result at 0.1$\,$Myr after the supernova explosion (ground truth). Panel (c): the 3D-MIM's forecast result at 0.1$\,$Myr. 
    Red crosses show the centre of the explosion. In panels (b) and (c), red dotted circles represent the shell radius estimated by the analytic solution (\ref{eq:Sedov}).
    Orange dashed lines show the boundary of the regions where the densities decrease by more than 10 per cent of the initial densities by the explosion.}
    One side of each panel corresponds to 60 pc.
    Colour maps show the density distribution.
    The colour bar and scale are the same in panels (a)--(c).
    \label{fig:IPNonUni}
\end{figure*}

We show the test results on our 3D-MIM using four different test data (Table \ref{tab:experiments_table}).
We first use the Fiducial dataset for the training and test dataset (Experiment 1; Table \ref{tab:experiments_table}). The training and test datasets are generated from different sets of simulations (random seeds).
Fig. \ref{fig:IPNonUni} shows an example of the forecast result using 3D-MIM.
Panels (a), (b) and (c) of Fig.~\ref{fig:IPNonUni} show the initial distribution, the simulated distribution (ground truth) and the forecast result at $t = 20\Delta t = 0.13$ Myr, respectively. The red crosses in the centre represent the position of the SN explosion.

In panels (b) and (c) of Fig.~\ref{fig:IPNonUni}, 
we show the shell radius in orange dashed lines.
The shell is defined by the boundary of the regions where the density decreases by more than 10 per cent of the initial densities for simplicity.
For reference, the red dotted circles represent the shell radius estimated using the analytic solution given by equation (\ref{eq:Sedov}).
In the simulated result (Panel (b)), while the shell spreads out as much as in the analytic solution on the left, the blast wave is stopped by a dense filament on the lower right, making the spread of the shell smaller.
A similar trend is also observed in the predicted result (Panel (c)), showing that the 3D-MIM can properly reproduce the anisotropic shell propagation.
Additional examples of the prediction are laid on Appendix \ref{sec:addition}.

\begin{figure*}
	\includegraphics[width=2\columnwidth]{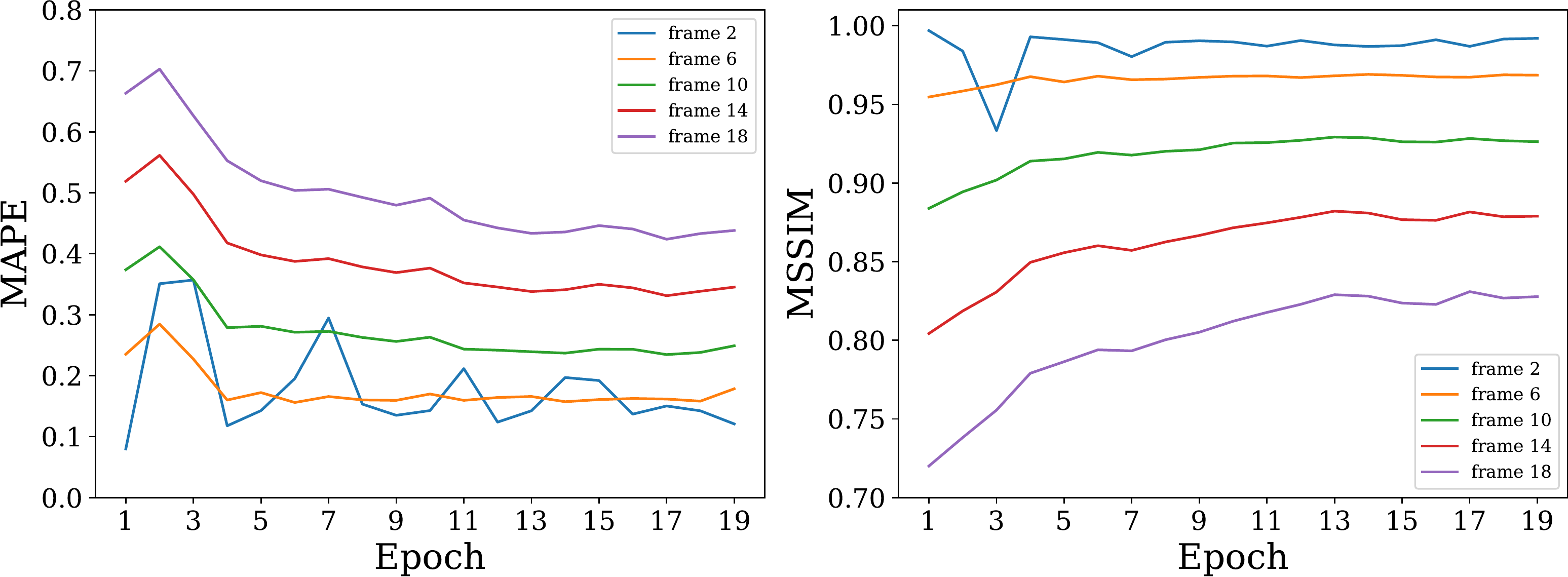}
    \caption{{\it Left}: mean absolute percentage error (MAPE). {\it Right}: mean structural similarity (MSSIM) as functions of the epoch at several frames for Experiment 1. Each colour corresponds to $t = 1.4 \times 10^{-2}, 4.2 \times 10^{-2}, 7.0 \times 10^{-2}, 9.8 \times 10^{-2}  ,{\rm and}~1.26 \times 10^{-1}$ Myr  (frame 2, 6, 10, 14, and 18) after the SN explosion. The lower value of MAPE and the higher value of MSSIM mean that the two images are more similar.
    }
    \label{fig:Eval}
\end{figure*}

\begin{figure*}
	\includegraphics[width=2\columnwidth]{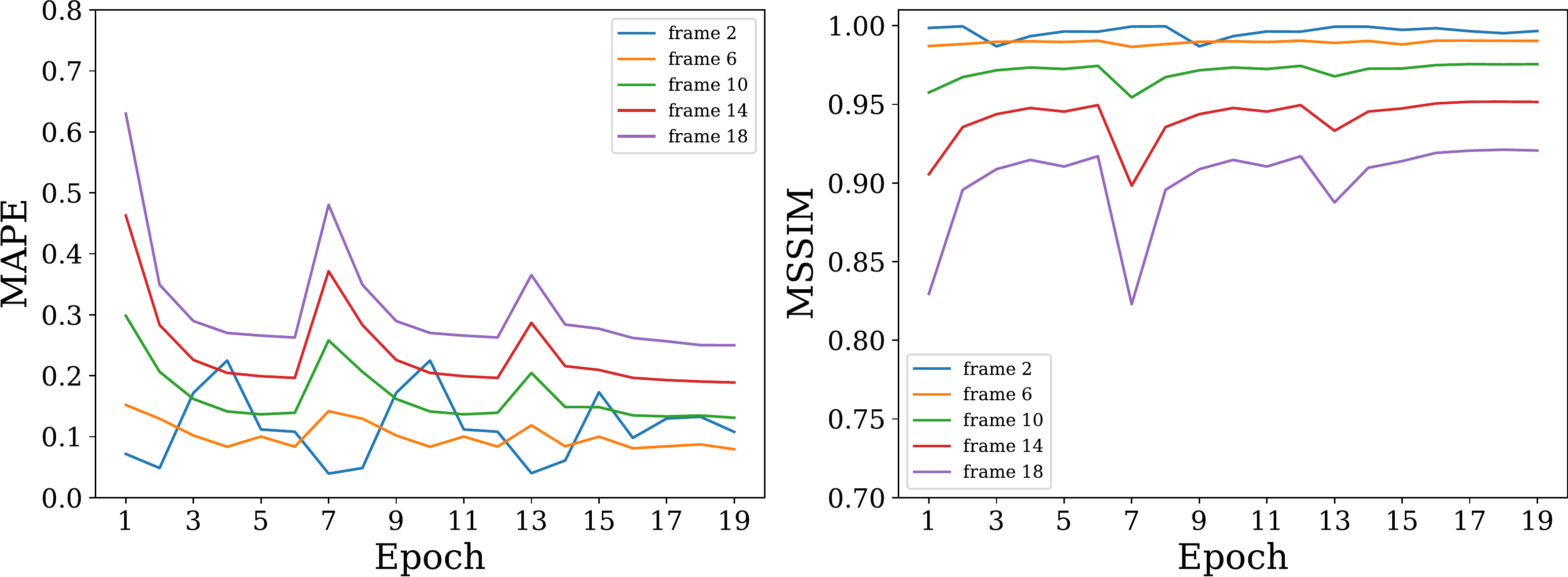}
    \caption{{\it Left}: mean absolute percentage error (MAPE). {\it Right}: mean structural similarity (MSSIM) as functions of the epoch at several frames for Experiment 4. Each colour corresponds to $t = 7.0 \times 10^{-3}, 2.1 \times 10^{-2}, 3.5 \times 10^{-2}, 4.9 \times 10^{-2},{\rm and}~6.1 \times 10^{-2}$ Myr  (frame 2, 6, 10, 14, and 18) after the SN explosion.
    }
    \label{fig:Eval_5e4}
\end{figure*}

To confirm the convergence of the training,
we compute two metrics: Mean Absolute Percentage Error (MAPE) and Mean Structural SIMilarity \citep[MSSIM;][]{Wang+2004} between the simulated and forecast data.
The MAPE measures prediction accuracy for every predicted voxels and ranges in $[0,1]$.
The MSSIM quantifies the structural similarity between simulated and predicted density distribution and ranges in $[-1,1]$. 
The lower value of MAPE and the higher value of MSSIM indicate that the two images are more similar.
See Appendix \ref{sec:index} for more details on these metrics.
MAPE and MSSIM are calculated by equations \ref{eq:mape} and \ref{eq:mssim}, respectively.

Fig. \ref{fig:Eval} shows these metrics as a function of epochs
for Experiment 1, where we use the Fiducial dataset for training and testing.
Each colour corresponds to a different frame.
We find that these metrics converge as the number of epochs increases.
This suggests that our model does not suffer from overfitting.
It also implies that the parameters' weights of the neural network also converge and that the model does not improve further with continued training.
Frame-by-frame comparisons of these two metrics show that the longer the forecast period is, the more difficult the reproduction becomes.
When we use other experiments, these metrics as functions of epochs show a similar tendency (Fig. \ref{fig:Eval_5e4}).

\begin{figure}
	\includegraphics[width=\columnwidth]{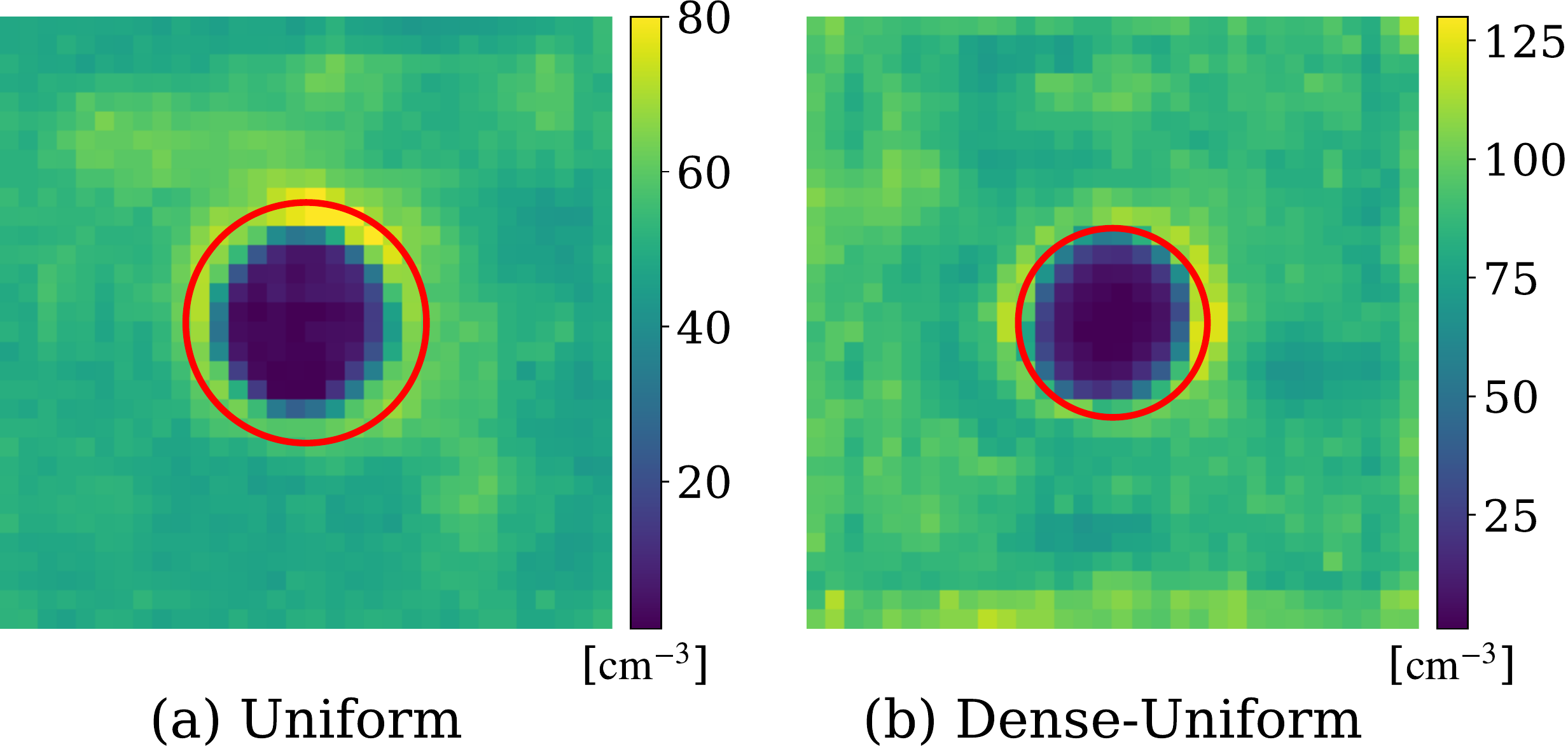}
    \caption{The test results of the shell expansion at 0.1$\,$Myr after a SN explosion with Experiment 2 (the panel (a)) and Experiment 3 (the panel (b)) in Table \ref{tab:experiments_table}.
    One side of each panel corresponds to 60 pc.
    The uniform densities of panels (a) and (b) are equal to and more than the mean density in SN simulations in training data, respectively.}
    We note that our model is not trained using the dataset of uniform density distributions but the Fiducial dataset in Table \ref{tab:datasets_table}.
    Red circles indicate the expanding shell radius obtained from the Sedov-Taylor solution (equation (\ref{eq:Sedov})).
    \label{fig:Const}
\end{figure}

To further examine the general performance of our model, the difference between the radius predicted by the model and the analytic solution using equation \ref{eq:Sedov} is tested.
In this test, we use both approaches to predict the radius of SN shells in uniform-density distributions.
We note that the model was not trained with such uniform-density SN explosion simulations.
Fig. \ref{fig:Const} shows the test results at 0.1 Myr of Experiment 2 (left) and Experiment 3 (right). 
In Experiment 2, we use the same average density as the training data ($5.6 \times 10$~cm$^{-3}$, left), whereas in Experiment 3, we adopt a higher density ($1.9 \times 10^2$~cm$^{-3}$, right).
In both cases, background gas is uniform and isotropic.
In Fig. \ref{fig:Const}, the red circles show the Sedov solution calculated using equation \ref{eq:Sedov}.
The shell (high-density peak) in the forecast result matches the analytic solution well when the uniform unit density is input (left), although the model is not trained with such a uniform distribution. 
When the ISM with high density is adopted (right), the estimated shell size is smaller than that of the ISM with mean density (left). Although the expected density dependence trend from the analytic solution is accurate, there are slight discrepancies in the sizes. This difference suggests that the model has not completely learned the scaling law.

To correctly predict it, we may need to vary the densities in training data more.
Another possibility is to train multiple 3D-MIMs with different density ranges and choose an appropriate one to apply every time a SN occurs in simulation. In practice, the average density of the ambient ISM must be calculated each time, but the computational cost for this is small enough.

\begin{table}
	\centering
	\caption{The settings for experiments.
	The names of datasets correspond to those in Table
	\ref{tab:datasets_table}. Experiments 1, 2, and 3 use the same trained models.
	}
	\label{tab:experiments_table}
	\begin{tabular}{lll} 
		\hline
		     Experiment & Training Data & Test Data \\
		\hline
		Experiment 1 & Fiducial & Fiducial\\
		Experiment 2 & Fiducial & Uniform\\
		Experiment 3 & Fiducial & Dense-Uniform\\
		Experiment 4 & Short-Term & Short-Term\\
		\hline
	\end{tabular}
\end{table}

\section{Discussion}
\label{sec:discussion}

\subsection{Application for the Identification of SN-Affected Particles}
\label{sec:particle-selection}

So far, we have seen that our model sufficiently predicts the expansion of SN shells.
Such a forecast result could be used for improving future galaxy simulations.

\begin{figure}
	\includegraphics[width=\columnwidth]{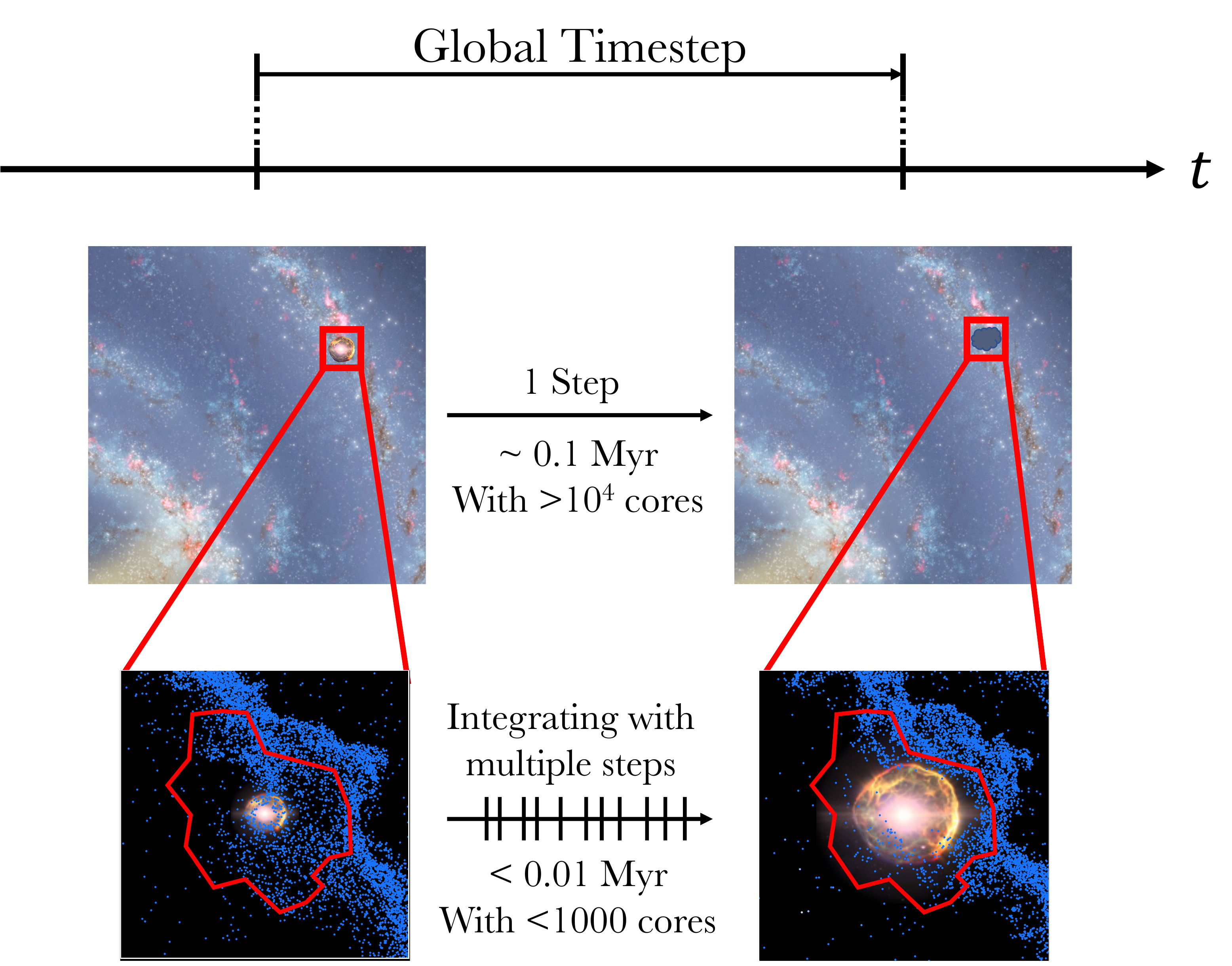}
    \caption{
    Upper panels show the galaxy's spiral arms in a simulation at two global timesteps, and lower panels display the zoom-in of the arm where an SN is exploding. 
    In the Hamiltonian splitting method, the particles affected by SNe are integrated separately from the entire galaxy with a short timestep and a low parallelization degree. 
    To apply this method, we need to predict the region 
    where the particles are affected by SNe during the global timestep (i.e. the region surrounded by the red line) in advance.
    Image credit:[{\it upper}: NASA/JPL-Caltech/ESO/R. Hurt, {\it lower}: ESA/Hubble (L. Calçada).]}
    \label{fig:Bridge}
\end{figure}

As discussed in Section \ref{sec:intro}, one way to address the issues of communication overhead in galaxy simulations is a Hamiltonian splitting method.
In this method, the Hamiltonian of a system is split into two parts: short and long timescales. Particles with short timescales (e.g. SN-affected particles) are integrated with short timesteps, and the force from long-timescale particles is given as a perturbation in a longer timestep (global timestep) \citep{Wisdom+1991}.
For future high-resolution galaxy simulations, we consider splitting the Hamiltonian into terms for the SN-affected particles and the others as illustrated in Fig. \ref{fig:Bridge}.
At every global timestep, we separately integrate the SN-affected particles using short timesteps shown in lower panels in Fig. \ref{fig:Bridge}.
Although those particles are challenging to analytically detect in advance, they could be selected using a forecast result by our deep-learning model.

We define {\it target} particles as those that require smaller timesteps than the subsequent global timestep and have a temperature above 100 K. The temperature threshold is given to exclude star-forming gas.
We call the rest of the particles {\it non-target} particles.
In our deep learning model, the central bubbles can be detected by extracting the regions where the density decreases at a certain level. As a demonstration, we consider regions with more than 10 per cent density reduction.
By image processing, we then slightly expand the detected bubble to cover the surrounding shell that also holds target particles (see Appendix \ref{sec:CV}).

In comparison to the analytic approach detailed in Appendix \ref{sec:analytic}, which relies on equation (\ref{eq:Sedov}), we assess the computer vision approach. 
We compute the {\it target} identification rate (the ratio of identified particles to true {\it target} particles) and {\it non-target} fraction (the ratio of {\it non-target} to {\it target} particles inside the forecast region).
The parameters for particle identification are chosen so that we obtain similar {\it non-target} fractions for the analytic solution (\ref{eq:Sedov}) and our 3D-MIM. 

\begin{figure}
	\includegraphics[width=\columnwidth]{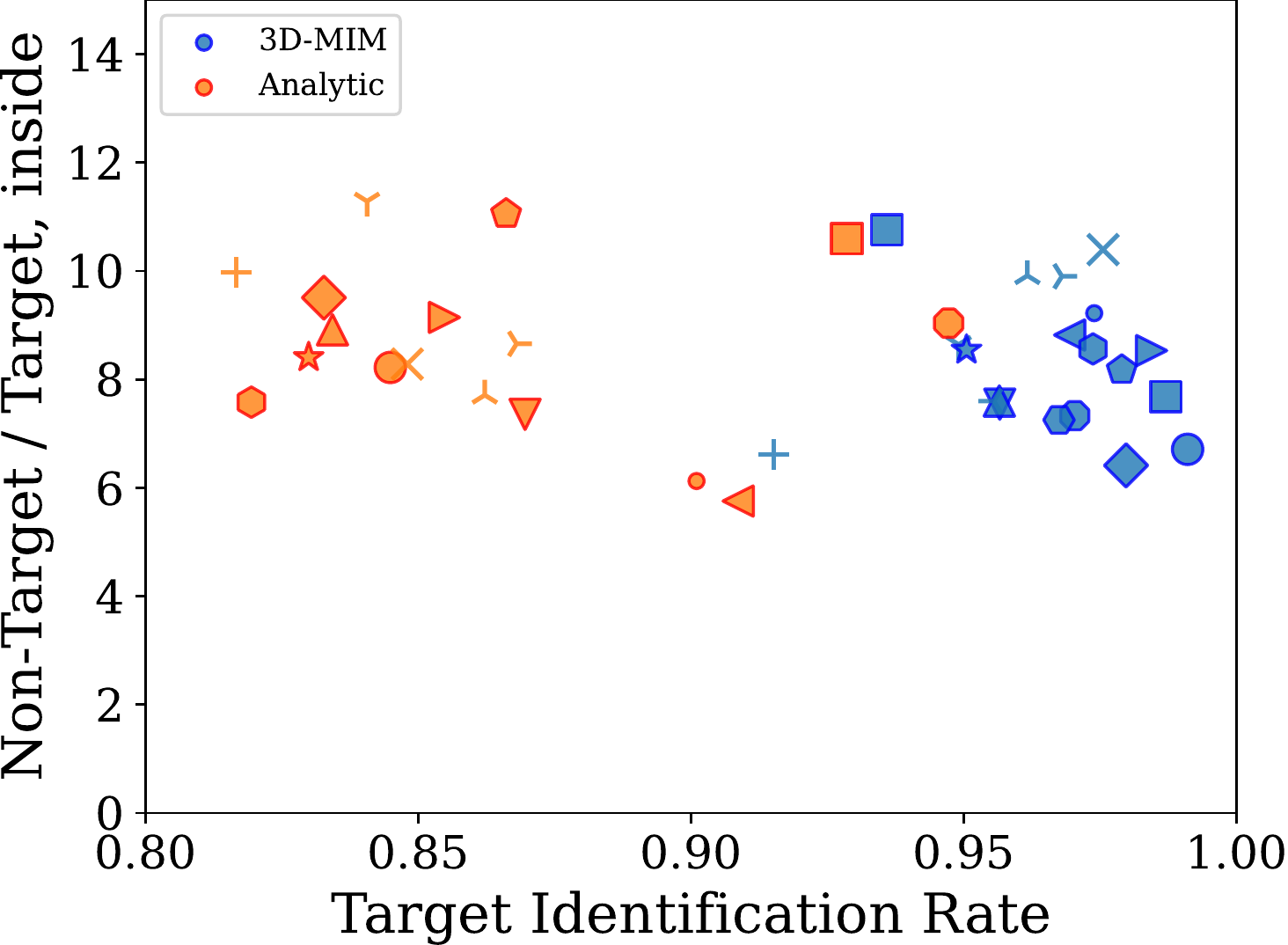} 
    \caption{The comparison of the approach based on the analytic solution (\ref{eq:Sedov}) (represented as ``Analytic'') and our model (represented as ``3D-MIM'') for global timesteps of $t=0.1$ Myr. The vertical axis indicates the ratio of the {\it non-target} to the {\it target} particles inside the forecast region. The horizontal axis indicates the fraction of {\it target} particles we successfully detected.
    Pairs of the same symbols indicate the results using the same simulation data.
    }
    \label{fig:Comp1Msun}
\end{figure}

Fig. \ref{fig:Comp1Msun} shows the results of our approach (blue) and the analytic approach (orange) for Experiment 1 (Table \ref{tab:experiments_table}).
The identification rates of our approach are significantly higher than those of the analytic approach.
This is because our deep learning model can capture anisotropically distributed particles long after the explosion, whereas the analytic solutions cannot. 

In general, there is a trade-off; a very high identification rate could be obtained if we allow a high {\it non-target} fraction.
In actual simulations, the parameters (e.g. criteria) for particle detection must be selected according to the required selection efficiency. In particular, we should adopt an identification rate with which the reproduction of the blast wave can be achieved within the required accuracy by simply evolving only the selected particles. 

\subsection{Application to Galaxy Simulations}
\label{sec:OtherPhysics}
This section discusses other physical processes that we should consider in the practical implementation of our model in future galaxy simulations. 

Several events other than SNe would also reduce the required calculation timesteps in cosmological simulations of the MW-sized galaxy. 
For instance, merging the main haloes and ram pressure-stripped satellites can generate shock-heated gas that reduces timesteps. However, their timesteps would not be as small as those for SNe because the Galactic halo gas is much less dense than that of disk region \citep[$10^{-5} - 10^{-3} ~\mathrm{cm^{-3}}$; ][]{Bland-Hawthorn+16} compared to star-forming regions, which we focus on in this paper.

When we apply the Hamiltonian splitting, we should also confirm that the physical processes other than SN feedback that affect ISM can be incorporated properly.
One of the most important effects is the far ultraviolet (FUV) radiation, which impacts cooling and heating \citep{Hu+2017, Hu+2019, Hislop+2022}.
The heating rate from FUV photons is proportional to their flux in cold and dense regions such as star-forming regions \citep{Forbes+16}.
The energy density of a gas particle contributed by (far-) UV luminosity $u$ is calculated as
\begin{equation}
    u=\sum_{i} \frac{L_i}{4 \pi c r_i ^2},
    \label{eq:luminosity}
\end{equation}
where $c$, $L_i$, and $r_i$ represent the speed of light, (far-) UV luminosity of a star particle $i$, and distance between the gas and star particle, respectively \citep{Hu+2017}.
For the contribution from outside the split region (SN-affected region), the radiation can be considered not to change quickly, and thus the summation in equation (\ref{eq:luminosity}) can be executed only at every global timestep, i.e., at the same time as the gravity tree walking.
For the internal radiation, if any, as the self-shielding is expected to be effective inside dense gas clouds, it would be sufficient to consider that it only contributes locally, similar to the treatment in \citet{Hu+2017}.
This way, other physical processes can still be computed adequately with Hamiltonian splitting methods.

\section{Conclusion}
\label{sec:conclusion}


We developed a new deep learning model, 3D-MIM, successfully forecasting the time evolution of the gas density distribution for 0.1\,Myr after a SN explosion.
To apply for 3D numerical simulations, the model is expanded from a 2D model originally utilized by video prediction by increasing the dimensionality.
Compared to the analytic solution, our method has the advantage of predicting the evolution of non-uniform gas clouds.
The performance of the 3D-MIM was evaluated using the metrics of MAPE and MSSIM for image reproductions. Additionally, we forecast the shell expansion of a SN in a uniform density distribution, which was not included in the training dataset. The 3D-MIM successfully reproduced the shell radius consistent with an analytic solution and demonstrated high convergence values and generalization capabilities.

Such a method presented in this paper could be used to pre-identify the SN particles that require short timesteps in large, high-resolution galaxy formation simulations. 
By combining this with the Hamiltonian splitting method, we would be able to integrate such particles separately from the entire galaxy.
In the future study, we will include our deep-learning model and Hamiltonian Splitting method in our $N$-body/SPH code, \textsc{ASURA-FDPS} \citep{Saitoh+2008,Iwasawa+2016}, to achieve a star-by-star galaxy formation simulation.
We note that some practical issues still need to be discussed in more detail, including the required identification rate and the implementation of other physical processes. We will investigate them in the future study.

Finally, as an interesting application of our 3D-MIM, we briefly mention the possibility of replacing the time-consuming computation of SNe with machine prediction, which would require much smaller computational costs than adopting the splitting scheme we consider in this paper. 
We can develop such a machine by designing it to learn the distributions of all the physical quantities, including velocity and temperature.
Such an attempt at replacement has been actively studied in recent years \citep[][]{Duarte+2022, Chan+2022},
but it has several technically challenging problems to overcome.
For example, an extreme amount of simulations for training data is necessary to obtain robust prediction over various physical conditions in a simulation.
It is also required to find appropriate transform functions for the training data to allow the machine to properly learn the physical quantities spread over several orders of magnitude \citep[e.g.][]{Chan+2022}.
We will also explore the future use of 3D-MIM in this direction.

\section*{Acknowledgements}
The authors thank an anonymous referee and the referee, Ulrich P. Steinwandel, for the useful comments on the paper.
The authors are also grateful to Takashi Okamoto for fruitful discussions.
Numerical computations were carried out on Cray
XC50 CPU-cluster at the Center for Computational Astrophysics (CfCA) of the National Astronomical Observatory of Japan.
The model was also trained using the FUJITSU Supercomputer PRIMEHPC FX1000 and FUJITSU Server PRIMERGY GX2570 (Wisteria/BDEC-01) at the Information Technology Center, The University of Tokyo.
This work was also supported by JSPS KAKENHI Grant Numbers 22H01259, 22KJ0157, 20K14532, 21H04499, 21K03614, and 23K03446, and MEXT as “Program for Promoting Researches on the Supercomputer Fugaku” (Structure and Evolution of the Universe Unraveled by Fusion of Simulation and AI; Grant Number JPMXP1020230406).
K.H. is financially supported by JSPS Research Fellowship for Young Scientists and accompanying Grants-in-Aid for JSPS Fellows (22J23077), JEES $\cdot$ Mistubishi corporation science technology student scholarship in 2022, and the IIW program of The University of Tokyo.

\section*{Data Availability}
The \textsc{3D-MIM} is open-source on GitHub at \url{https://github.com/kyaFUK/3D-MIM}. The data supporting this study's findings are available from the corresponding author, K.H., upon reasonable request.



\bibliographystyle{mnras}
\bibliography{example} 






\appendix

\section{Resolution Study of Blastwave test}
\label{sec:blastwave}

We test the Sedov explosion problem \citep[][]{Sedov1959} using our DISPH code \citep[][]{Saitoh+2013,Saitoh+2016}.
The parameters for this test are similar to the point-like explosion tests carried out in \citet[][]{Saitoh+2009}.
Since this is an idealized test, unlike Section \ref{sec:code}, we do not consider self-gravity and radiative cooling/heating.
The equations we use in this test are as follows:
\begin{align}
    P&=(\gamma-1) \rho u,\\
    \frac{d \rho}{dt} & = -\rho \nabla \cdot \bm{v}, \\
    \frac{d^2 \bm{r}}{dt^2} & = -\frac{\nabla P}{\rho} + \bm{a}_{\rm visc}, \\
    \frac{du}{dt} & = -\frac{P}{\rho} \nabla \cdot \bm{v}.
\end{align}

As described in Section \ref{sec:method_analytic}, the system of a spherical point explosion in a uniform media has the dimensionless similarity variable of the Sedov solution.
By transforming equation (\ref{eq:Sedov}), the dimensionless similarity variable $\xi$ can be written as
\begin{equation}
    \xi = \left( \frac{\rho_0}{E_0} \right)\frac{r}{t^{2/5}},
    \label{eq:xi}
\end{equation}
where $E_0$, $\rho_0$, $r$, and $t$ is the injected energy, initial density, radius of the shell, and elapsed time, respectively.
These physical quantities are normalized to the dimensionless parameter in this test: $\rho_0$ and $E_0$ are normalized to unity.
A SN explodes at $r=0$ at $t=0$, and $64^3$ particles are used.

\begin{figure}
	\includegraphics[width=\columnwidth]{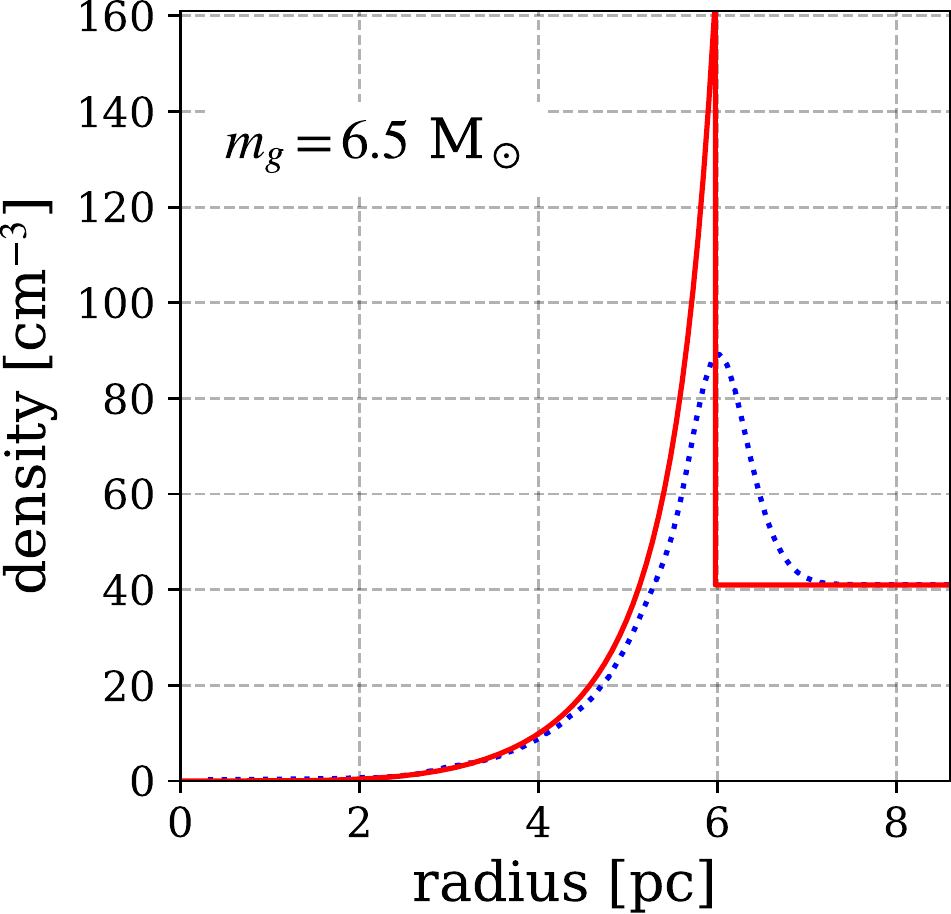} 
    \caption{ The simulation result of the Sedov explosion problem \citep[][]{Sedov1959}. 
    The simulation was performed with the DISPH scheme \citep[][]{Saitoh+2013,Saitoh+2016} that is used in our SN simulation code. The result is scaled to the mean density of $4.1 \times 10 ~{\rm cm^{-3}}$ at $t =10^4$~yr. {\it Red solid curve}: the analytic solution for the radial density profile of the blast wave derived by \citet{Sedov1959}. {\it Blue dotted curve}: geometrical means of the density in the simulation.
    }
    \label{fig:BW}
\end{figure}
Fig. \ref{fig:BW} shows the simulation result at $t=5.0 \times 10^{-2}$ for the dimensionless time.
Applying the dimensionless parameters to equation (\ref{eq:xi}), the value of the dimensionless similarity valuable $\xi=1.13$ is obtained while $\xi=1.15$ for the adiabatic gas analytically derived by \citet[][]{Sedov1959}.
Using $\xi=1.15$, the dimensionless parameters, and equation (\ref{eq:Sedov}), the obtained relationship between the radius and density are scaled to the similar parameters as the sparsest case in our training dataset: the mean density of the ambient gas $4.1 \times 10 ~{\rm cm^{-3}}$, input energy $10^{51}$ erg, and elapsed time $10^4$ yr (Fig. \ref{fig:BW}).
From this test, the required mass resolution is obtained as 6.5 ${\rm M_\odot}$.
The mass resolution $m_{\rm gas}=$ 1 M $_\odot$, ~chosen in our SN simulations for the training data, is sufficient.
This conclusion is also supported by \citet[][]{Steinwandel+2020}.
They also detected the blast wave under a similar condition using SPH with the mass resolution of 1 M$_\odot$ \citep[Table. 1 in ][]{Steinwandel+2020}.

We conduct an additional test on the SN feedback in a spherical gas cloud with a uniform density to confirm the convergence with the mass resolution.
Such tests have also been done in Appendix B in \citet{Hu+2016}, 
but here we adopt higher metallicity and include the gravitational forces to evaluate the convergence in more practical conditions.
We use three different resolutions ($m_{\rm gas}=$ 10 M$_\odot$, 1 M$_\odot$, and 0.1 M$_\odot$).
The injection masses are $M_{\rm inj} =$ 1000 M$_\odot$, 100 M$_\odot$, and 10 M$_\odot$, respectively.
We simulate the evolution of a supernova remnant (SNR) with the injected energy $E_{\rm SN} = 10^{51}$ erg in a spherical gas cloud with a uniform density $n_{\rm H} = 45 ~{\rm cm^{-3}}$.
The ambient gas has the initial temperature $T=100$ K.
We refer to the mass of the hot bubble ($T\geq 2 \cdot 10^4$ K) and cold shells ($T <2 \cdot 10^4$ K) to evaluate the convergence depending on the resolution following the definition in \citet{Steinwandel+2020}.
The threshold of velocity $v > 0.1 \mathrm{km s^{-1}}$ is also set to distinguish the shell and ambient gas.
The gravitational softening length for gas is set to be 2 pc, 0.9 pc, and 0.5 pc for the SPH particle mass of 10 M$_{\odot}$, 1 M$_{\odot}$, and 0.1 M$_{\odot}$.

\begin{figure*}
	\includegraphics[width=2\columnwidth]{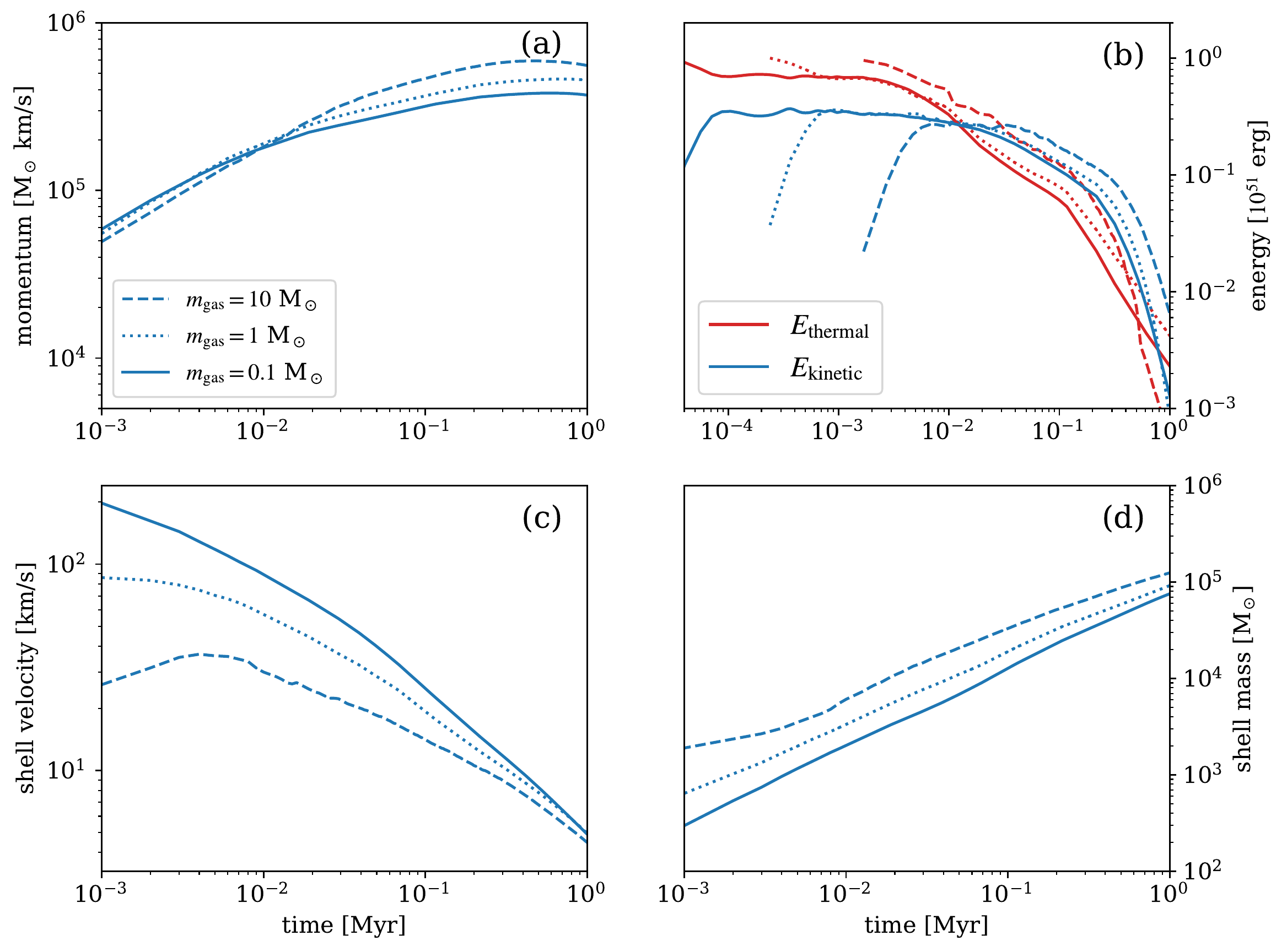}
    \caption{Evolution of a SN remnant in the cold neutral medium (the initial temperature $T=100$ K and density $n_0=$ 45 ~${\rm cm^{-3}}$). Three different mass resolutions are adopted: $m_{\rm gas}=0.1$ (solid), 1 (dotted), and 10 (dashed) M$_\odot$. Panel (a): the linear momentum of the shell; panel (b) the thermal energy (red) and kinetic energy (blue); panel (c) the shell velocity, defined by the total momentum divided by the shell mass; panel (d): the shell mass. The shell was defined by all particles, which have the temperature of $T < 2 \cdot 10^4$ K and the velocity $v>0.1 {\rm km s^{-1}}$.
    }
    \label{fig:convergence}
\end{figure*}

\begin{figure}
	\includegraphics[width=\columnwidth]{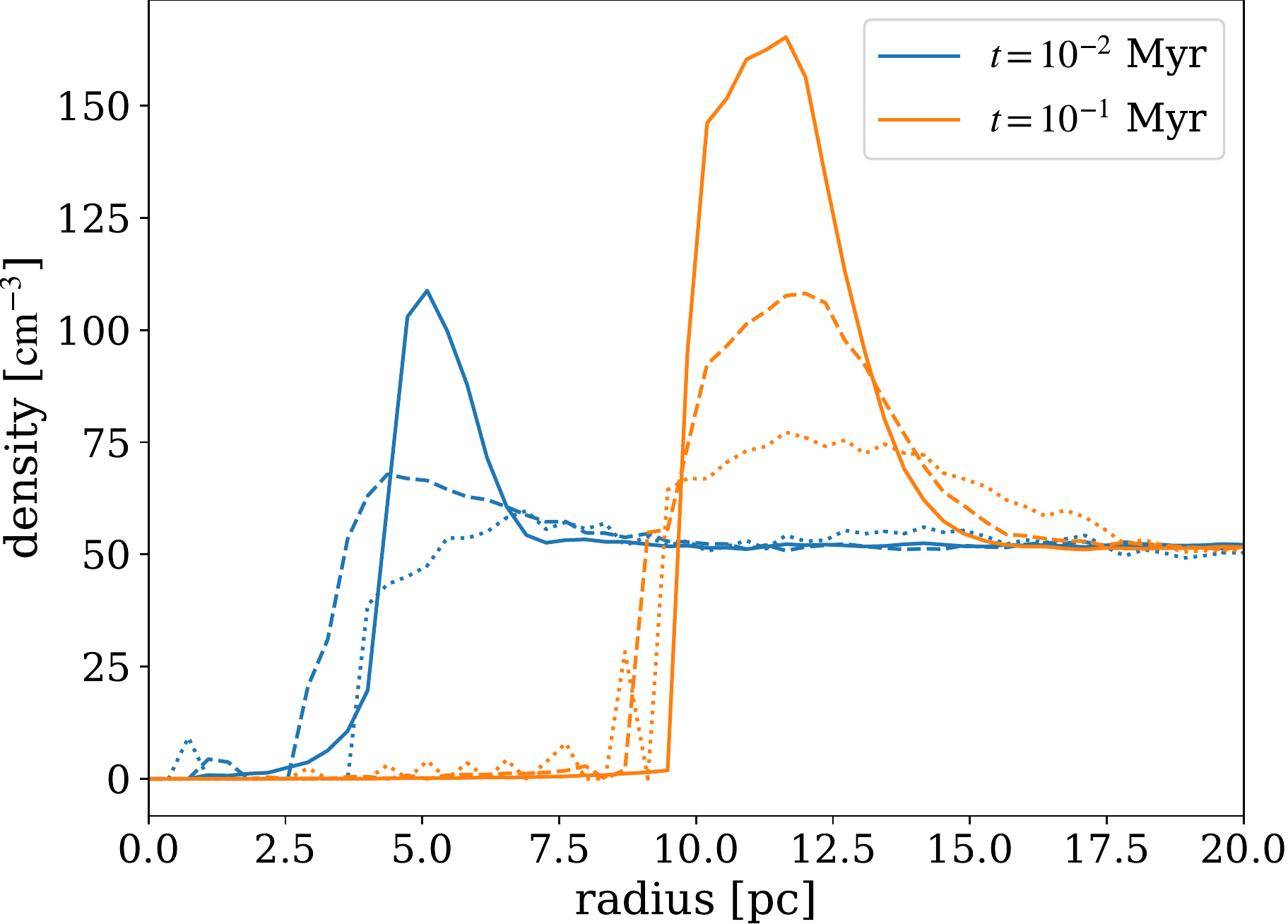}
    \caption{Density profile corresponding to the radius at the elapsed time at $t=10^{-2}$ Myr and $t=10^{-1}$ Myr from the energy injection. Line style means the same as Fig. \ref{fig:convergence}.
    }
    \label{fig:dens_prof}
\end{figure}

Fig \ref{fig:convergence} shows the evolution of the SN feedback.
The momentum is overestimated in panel (a) at the worst resolution (10 M$_\odot$). This is consistent with the results of  \citet{Steinwandel+2020} although they consider lower-metallicity environments.
Since the run with 10 M$_\odot$ resolution does not resolve the thin and high-density shell (Fig. \ref{fig:dens_prof}), the cooling timescale becomes much longer than the others. As a result, the shell holds the thermal energy for a longer time (see Fig. \ref{fig:convergence} (b)), and later it is converted into high momentum.
With these results, we chose 1 M$_\odot$ resolution, which is fine enough to reproduce the reasonably converged evolution of the SN shells.

\section{Metrics for image quality}
\label{sec:index}

This study uses two metrics, MAPE and MSSIM, to evaluate our deep learning model.
MAPE measures average reproducibility across the entire image,
while MSSIM measures local correlations between images.
Below, $\textbf{x}=\{x_i |\, i \in \mathbb{N} \}$ and $\textbf{y}=\{y_i |\, i \in \mathbb{N} \}$ denote the densities of the ground truth (simulation result) and predicted result, respectively.

\subsection{MAPE}

Mean Absolute Percentage Error (MAPE) is defined as
\begin{equation}
    \mbox{MAPE}(\textbf{x},\textbf{y}) = \frac{1}{N} \sum_{i=1} ^{N} \left| \frac{y_i - x_i}{x_i} \right|, 
    \label{eq:mape}
\end{equation}
where $N$ is the total number of voxels.

\subsection{MSSIM}
Structural SIMilarity (SSIM) is a metric for quality assessment based on the degradation of structural information \citep{Wang+2004}.
It is commonly used for the evaluation of image compression and restoration.
SSIM is defined as
\begin{equation}
    \mbox{SSIM}(\textbf{x},\textbf{y}) = \frac{(2\mu_x \mu_y +C_1)(2 \sigma_{xy} + C_2)}{(\mu_x^2 +\mu_y^2 +C_1)(\sigma_x^2 + \sigma_y^2 +C_2)}.
    \label{eq:mssim}
\end{equation}
where $\mu_x$ and $\mu_y$ ($\sigma_x$ and $\sigma_y$) are the mean (variance) of $\textbf{x}$ and $\textbf{y}$, respectively, and $\sigma_{xy}$ is the covariance of $\textbf{x}$ and $\textbf{y}$. 
$C_1$ and $C_2$ are small constants to avoid instability when the denominator is near zero.
The higher value of SSIM means that the two images are more similar.
The voxel-by-voxel match of the two images increases the value of $\sigma_{xy}$ and SSIM rather than the match on average.

In this study, we adopt $C_1=1.3 \times 10^{-7}$ and $C_2=1.2\times 10^{-6}$.
In general, SSIM is calculated at each local window with a given size \citep[e.g.][]{Dutta+2019,Zeng+2012} and the average of them is called Mean SSIM (MSSIM).
The higher value of MSSIM also means that the two images are more similar.
We calculate a local SSIM on each $8\times 8\times 8$ block window.
We divide a 3D volume image with the size of $32\times 32\times 32$ into $256(=4^3)$ blocks with the size of $8\times 8\times 8$
and take the MSSIM, i.e. the average SSIMs of the divided blocks.

\section{Example Results of the Prediction}
\label{sec:addition}
\begin{figure*}
 \includegraphics[width=1.9\columnwidth]{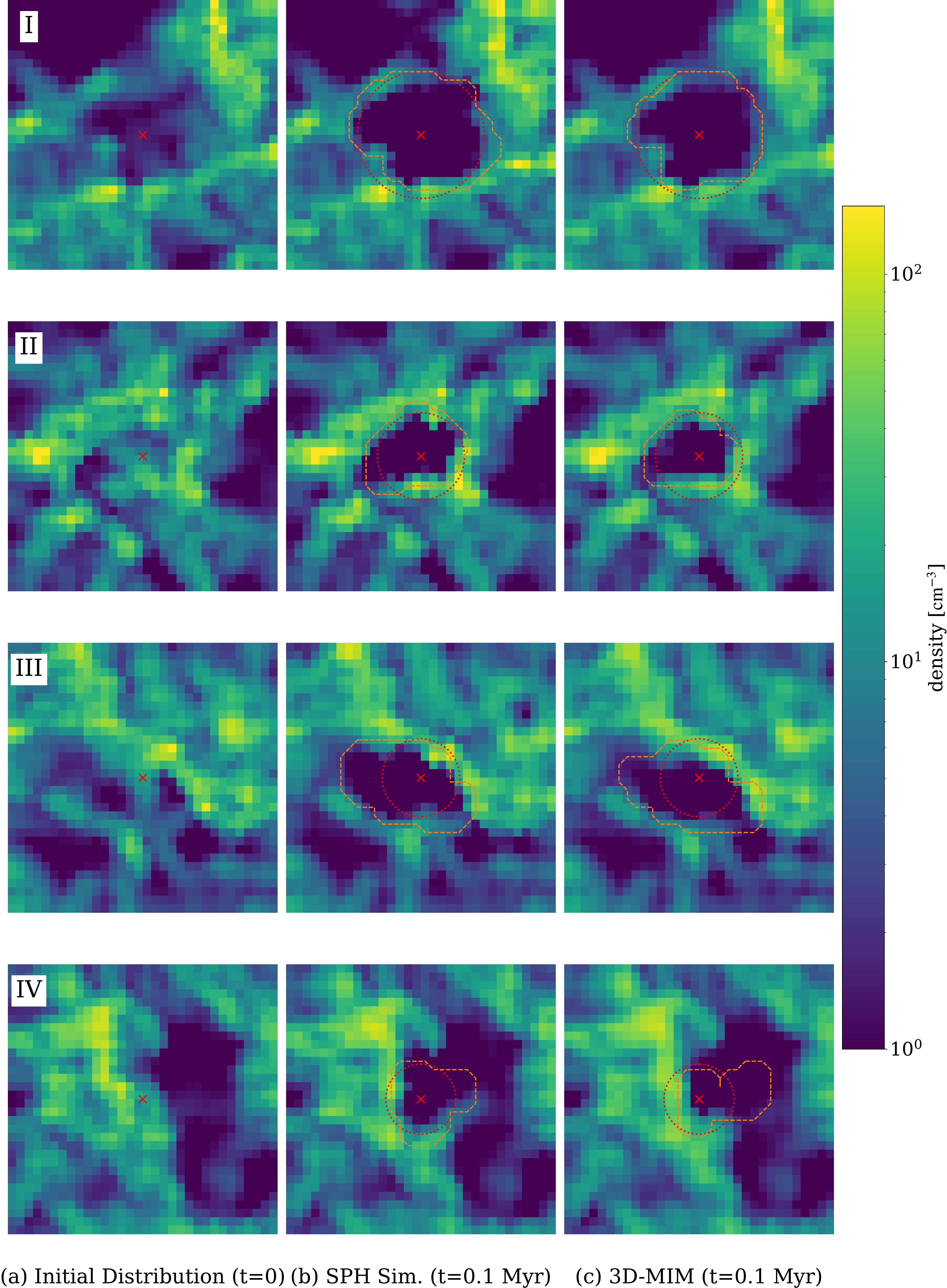}
    \caption{The 3D-MIM's prediction of the shell expansion in four different initial distributions.
    Panel I, II, III, and IV represent the prediction of those with the initial mean density in the region within 5 pc around the explosion centre, which are $3.2 \times 10$, $2.1 \times 10^2$, $3.7 \times 10^2$, and $6.3 \times 10^2$ ${\rm cm^{-3}}$, respectively.
    The symbols and lines represent the same as Fig. \ref{fig:IPNonUni}.
    }
    \label{fig:IPNonUni_app}
\end{figure*}

In Fig. \ref{fig:IPNonUni_app}, we present four examples of the prediction with the 3D-MIM in order to show that our prediction does not strongly depend on the initial gas distribution.
In particular, we show the predictions of the initial distribution with high density where we need to integrate particles with significantly short timesteps.
The left, central, and right columns show the initial distribution, the result of SPH simulations, and the prediction by the model, respectively.
Panel I, II, III, and IV represent the predictions of the shell expansion with the initial mean density in the region within 5 pc around the explosion centre, which are $3.2 \times 10$, $2.1 \times 10^2$, $3.7 \times 10^2$, and $6.3 \times 10^2$ ${\rm cm^{-3}}$, respectively.
Panel I, which has a relatively low density at the centre, shows a relatively spherical shell.
In panel II, a SN explodes surrounded by dense filaments, which prevent the blastwave from propagating.
In panels III and IV, a SN occurs within a dense filament. The blastwave breaks the filament and propagates farther in the sparse region.
We confirmed that our model can predict the inhomogeneous expansion of the shell in various densities.

\section{Approaches for Particle Selection}
\label{sec:approach}

\subsection{Computer Vision Approach}
\label{sec:CV}
\begin{figure}
	\includegraphics[width=\columnwidth]{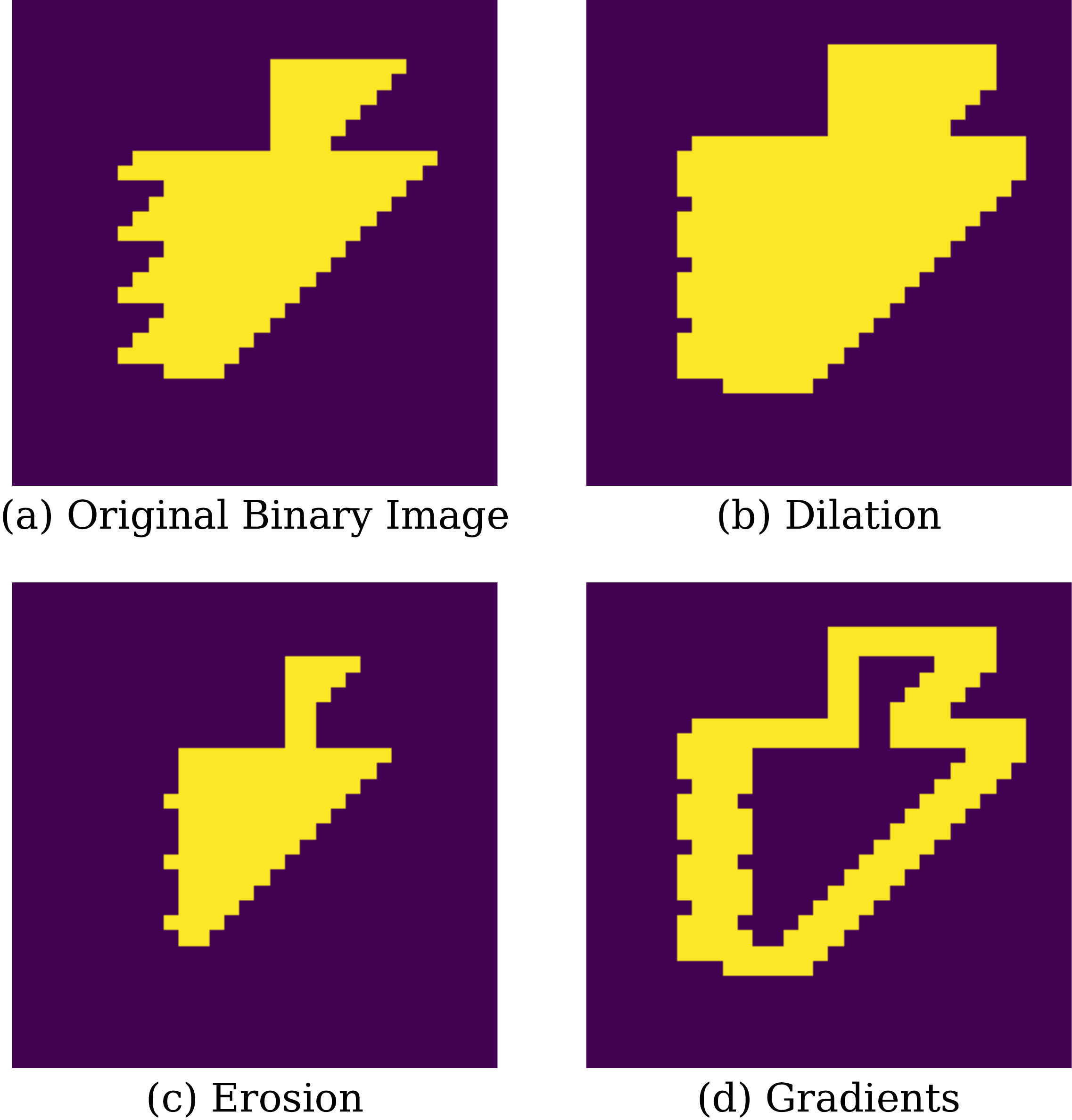}
    \caption{(a) Original binary image before the operations. (b)-(d) Results of three mathematical morphological operations used in our image processing for the original binary image.
    Each operation is done with a convolutional filter of the size of (3, 3).}
    \label{fig:MM}
\end{figure}

To extract a region with a smooth boundary and achieve a high identification rate, we employ three types of commonly used operators for image processing: {\it Dilation}, {\it Erosion}, and {\it Gradient}  \citep[ for review, ][]{Gonzalez+2006}.
An example of the operations is shown in Fig. \ref{fig:MM}.
The operators are applied for binarized images with a value of 0 or 1 assigned to each pixel, where we regard the set of pixels assigned 1 as the selected region.
The operators assign a value of 0 or 1 to a pixel based on the values of its neighbouring pixels.
By {\it Dilation}, 1 is assigned to a pixel when one or more neighbouring pixels have a value of 1.
Conversely, {\it Erosion} assigns 0 to a pixel when one or more neighbouring pixels have a value of 0.
{\it Gradient} compares the results of {\it Dilation} and {\it Erosion} and assigns 1 to pixels assigned different values by them.
The filter size determines the number of neighbouring pixels used in each operation.
In general, the {\it Dilation} operation enlarges the selected region by the filter size. 
{\it Erosion} and {\it Gradient} are used to round off the spiky boundaries of the selected region. 
{\it Erosion} gives a more subtle change than {\it Gradient} and is suited for making final adjustments later in the process.

Table \ref{tab:parameter_table_search} shows the order of the operations we adopt in this study.
{\it Erosion} and {\it Gradient} are executed to make the boundaries of the selected regions smooth. 
We adopt a filter size of 3 (minimum value) as the default value to preserve the region's shape as much as possible.
We also perform {\it Dilation} with a larger filter size as we find that it is needed to fill the inner cavity created by {\it Gradient}. 
Each processing is performed on 2D slices of the 3D volume image for three axes separately.
After a few 2D operations, we sum up the values of the corresponding voxels in the resulting three-volume images and then assign 1 to the voxels with a value above a threshold.
This operation ({\it thresholding}) is done twice during the processing.

\begin{table}
	\centering
	\caption{The order of operators applied to the images. 
	We use the operators of digital image processing, {\it Dilation}, {\it Erosion}, {\it Gradient}, and {\it Thresholding} \citep[][]{Gonzalez+2006}. 
	The filter sizes and the thresholds are listed in the second column. 
    }
	\label{tab:parameter_table_search}
	\begin{tabular}{lcc} 
		\hline
		Operators & Filter size / Threshold & \# of iteratios\\
		\hline \hline
		{\it Dilation} & (3,3) &  1\\
		{\it Gradient} & (3,3) &  1\\
		{\it Dilation} & (3,3) &  2\\
		{\it Dilation} & (5,5) &  2\\
		\hline
		{\it Thresholding} & $\geq 2$  & -\\
		\hline \hline
		{\it Erosion} & (3,3) & 1\\
		{\it Dilation} & (3,3)  & 2\\
		\hline
		{\it Thresholding} & $\geq 2$ & -\\
		\hline
	\end{tabular}
\end{table}

\subsection{Analytic approach}
\label{sec:analytic}
\begin{figure}
	\includegraphics[width=\columnwidth]{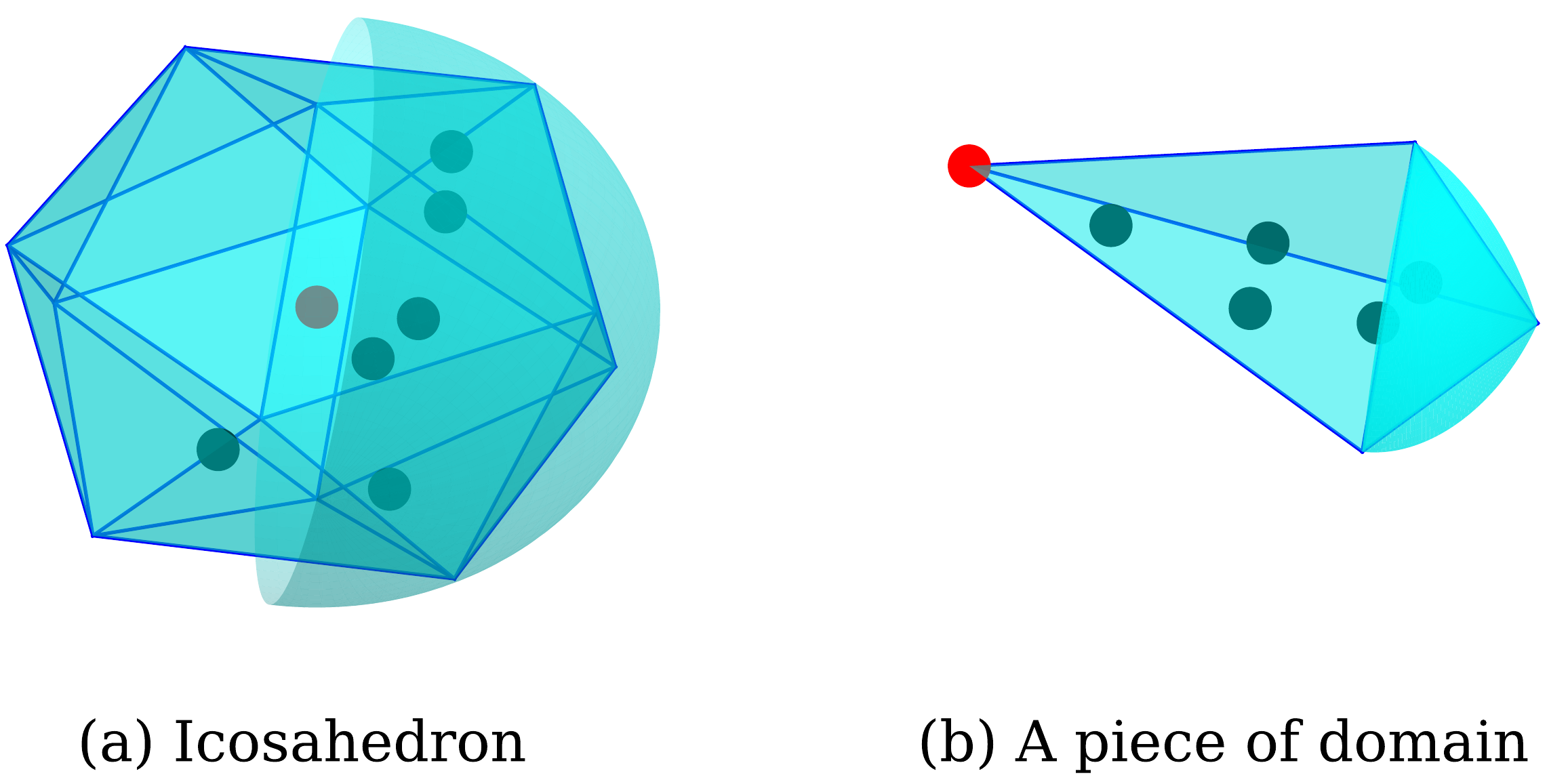}
    \caption{Domain decomposition using icosahedron (a) for calculating the shell expansion. We calculate the average density of the 20 tetrahedral cones (b) formed by connecting each vertex of the icosahedron to the centre and calculate the radius of the shell at a particular time in each direction using the 20 average densities and equation (\ref{eq:Sedov}).}
    \label{fig:DomainSp}
\end{figure}

Equation (\ref{eq:Sedov}) analytically predicts the radius of the shells under the ideal environment of uniform density. However, the interstellar gas is not uniform, and therefore this equation is not always applicable for predicting the evolution of SN shells.
One way to improve the analytic approach is by dividing the ambient gas into several regions along the radial direction of the SN.
Then we applied the analytic solution (equation \ref{eq:Sedov}) to the mean densities in each region.
This approximation is similar to the blast wave evolution model in \citet[][]{Haid+2016}.
We used an icosahedron domain decomposition, as shown in Fig.~\ref{fig:DomainSp}.
First, the centre of the icosahedron is set to the centre of the explosion (Fig.~\ref{fig:DomainSp} (a)). Second, we calculate the mean densities of  20 tetrahedral cones (Fig.~\ref{fig:DomainSp} (b)) formed by connecting each vertex of the icosahedron to the centre. 
Third, for each cone, the radius of the expanded shell is calculated with each density using equation (\ref{eq:Sedov}).
Then, we identify particles inside the radius of each shell as the particles that will require small timesteps.

\bsp	
\label{lastpage}
\end{document}